\documentstyle[12pt]{article}
\begin{document}
\pagestyle{myheadings} 
\thispagestyle{empty}
\vspace{3cm}

\begin{center}
{\bf STUDYING LEPTON FAMILY VIOLATION\\
     IN LEPTON-LEPTON COLLISIONS}
\end{center}
\vspace{1cm}

\begin{center}
 V.~V.~Kabachenko and Yu.~F.~Pirogov\\
\vspace{0.5cm}

{\it Theory Department, Institute for High Energy Physics,\\
Protvino, Moscow Region, RUSSIA\\
and\\
Moscow Institute of Physics and Technology,\\
Dolgoprudny, Moscow Region, RUSSIA}
\end{center}
\vspace{1cm}

\begin{abstract}
In the context of the future high energy -- high luminosity electron and
muon colliders, all the relevant four-lepton processes with the lepton family
violation (LFV) are systematically classified. The most general LFV effective
lagrangians are found, and the helicity differential cross sections for the
LFV processes are calculated. The six- and eight-lepton Standard Model 
(SM) backgrounds are discussed, and the LFV processes clean of the six-lepton 
background  are picked out. The possibility to suppress the six-lepton SM 
background, when present,  by the unnatural initial beam polarizations is 
investigated. 
It is shown that the four-lepton LFV processes are amenable to experimental 
study in the lepton-lepton collisions in the most favourable cases up to 
the underlying scale of order $10^2\,$TeV.
Studying these processes should  provide an essential part of the physics 
program for the next generation lepton colliders to unravel 
the outstanding flavour/family problem.
\end{abstract}

\newpage
\addtocounter{page}{-1}
\paragraph{Introduction}

At present there are two principle to be solved problems in the Standard Model
(SM): ({\em i}) the spontaneous symmetry breaking problem, or that of origin of
the electroweak gauge boson masses; ({\em ii}) the flavour/family problem, 
or that of origin of the fermion masses and mixings. 
The former problem is superficially
related with the TeV scale. It is expected to be mainly solved with the
near future advent of the LHC, followed by the discovery and subsequent study
of the Higgs boson. On the other hand, the latter problem should be related 
with a scale ${\cal O}(10$\,TeV) or higher  and would 
require for its study the multi-TeV energy facilities.

These facilities may be provided by the future high energy -- high luminosity
$e^+e^-$ and $e^-e^-$ linear colliders~\cite{ee}, or 
$\mu^+\mu^-$~\cite{mumu} and possibly $\mu^+\mu^+/\mu^-\mu^-$ ring colliders,
or conceivably even $\mu^+e^-$ and $\mu^-e^-$ ring-linac colliders. 
It is not our aim to discuss the feasibility of all these projects.
We would just like to emphasize that with the advent of such colliders a new 
era in the high energy physics, that of the unravelling the outstanding 
flavour/family problem,  might set in.

The present-day theories give no convincing explanation to this obscure
problem. So, an  experimental key to solving it may possibly  
be given by studying the processes 
with family violation, in particular the lepton family violation (LFV),
the latter being strictly forbidden in the SM. The
lepton-lepton colliders provide the unique clean environment for studying LFV.
We show that the advanced parameters of the colliders, 
namely $\sqrt{s}\ge 4$\,TeV and integrated luminosity $L\ge 10^3$\,fb$^{-1}$ 
conceivable at least for the $\mu\,\mu$ colliders, could allow one to study 
LFV in the four-lepton interactions up to the flavour/family scale 
${\cal O}(10^2$\,TeV) which is
quite assuring. We believe that this could provide an additional impetus in
favour of lepton colliders.

\paragraph{Four-lepton LFV interactions}

Whatever the underlying nature of the flavour/family might be, at the 
energies much 
lower then the corresponding mass scale it should most probably manifest itself 
in the family violating contact interactions, in particular in the 
four-lepton interactions, too.

The SM invariant four-fermion operators 
are generically of the type $\bar L\bar L LL$, $\bar R\bar RRR$, 
$\bar L\bar RLR$, $\bar L\bar LRR$ and $\bar R\bar RLL$. Here $L$ and $R$ 
denote the left-handed and right-handed lepton fields, respectively. 
In this, it can be shown that the last two operators are forbidden for 
the processes with leptons exclusively. It follows thereof
that all the allowed four-lepton operators can be reduced by the Fiertz 
rearrangement to the chirality conserving vector form
\begin{eqnarray}\label{operators}
{\cal O}_{LL}&=&\bar L\gamma_\mu L\,\bar L\gamma_\mu L, \nonumber\\
{\cal O}_{RR}&=&\bar R\gamma_\mu R\,\bar R\gamma_\mu R, \\
{\cal O}_{LR}&=&\bar L\gamma_\mu L\,\bar R\gamma_\mu R, \nonumber
\end{eqnarray}
which we will use in what follows\footnote{This form does not mean 
with necessity that all
the underlying exchanges are of the vector type. They can equally
well be transformed to the chirality violating scalar and tensor form. We use 
only vector type operators to fix somehow an ambiguity in the relative 
normalization of the different type operators.}. Each of the fields in 
Eq.~\ref{operators}
can represent the lepton of any family, what in fact proliferates the number
of independent operators. With this caveat in mind, the most general four-lepton
effective lagrangian can be written as
\begin{equation}\label{lagrangian}
{\cal L}=\frac{1}{F^2}\sum_i\frac{c_i}{1+\delta_i}{\cal O}_i,
\end{equation}
where $\delta_i=1$ for the operators with identical currents and $\delta_i=0$
for operators with different currents. Here $c_i$ are the 
generic free parameters, supposedly $c_i={\cal O}(1)$, and $F$ is an 
unknown flavour/family violation mass 
scale\footnote{We take for the effective coupling constant 
$g^2_{eff}=1$. In the usual notations with $g_{eff}^2=4\pi$, this would 
correspond to the scale $\Lambda=\sqrt{4\pi}F$.}. Our expectation is that 
$F={\cal O}(10^{2\pm1}$\,TeV).

There exist three possible types of the four-lepton LFV processes, 
namely $l'lll$, $l'l'll$ and $l''l'll$. Here $l$, $l'$ and $l''$ 
$(l\neq l'\neq l'')$ denote charged leptons/antileptons of different families. 
Let us consider these types of the LFV processes in turn.

\paragraph{(1) \boldmath  $l'lll$ type \unboldmath}

There are four independent operators in the effective lagrangian
\begin{equation}\label{l'lll}
{\cal L}=\frac{1}{F^2}\biggl(\eta_{LL}(\bar l'l)_L\,(\bar ll)_L
+\eta_{LR}(\bar l'l)_L\,(\bar ll)_R +(L\leftrightarrow R)+\mbox{h.c.}\biggr),
\end{equation}
where $(\bar ll)_L\equiv\bar l_L\gamma_\mu l_L$, etc. Here and in what follows 
all  free parameters $\eta$ etc are in general complex. 
Particular classes within the type, relevant for the electron and muon 
colliders, are $\mu eee$, $e\mu\mu\mu$, $\tau eee$ and $\tau\mu\mu\mu$.

\boldmath
{\bf ({\bf\em i}) $\mu eee$}
\unboldmath\ 
$\;$
Here one has $l'=\mu$, $l=e$, and the relevant processes in this case 
are\footnote{Here and later on we do not consider separately the  charge 
conjugate initial channels. The substitutions required 
for them are obvious.}                                  
\begin{eqnarray}\label{l'lll-processes}
&(s)&\;\;\;\;\;\cases{e^+e^-\to\mu^+e^-,\;e^+\mu^-,\cr
		    \mu^+e^-\to e^+e^-,\cr}\nonumber\\
&(u)&\;\;\;\;\;\cases{e^-e^-\to\mu^-e^-,\cr
                     \mu^-e^-\to e^-e^-.\cr}
\end{eqnarray}
Here and in what follows it will be chosen as the $s$ channel that one with 
zero net electric charge $Q=0$, and as $u$ channel the one with $|Q|=2$. 
The proper differential cross section for the $s$ channel with unpolarized 
initial particles is\footnote{For simplicity we use here and in what follows 
only the real free parameters. For complex parameters one needs 
the obvious replacements $\eta^2\to |\eta|^2$ and so on, 
the interference terms being absent in the limit $m_l\ll\sqrt{s}$. 
Hence the $CP$ violating LFV processes in the lepton-lepton collisions  
at the high energies are suppressed.}
\begin{equation}\label{section1}
8\pi s F^4\frac{d\sigma}{d\cos\theta}=(\eta_{LL}^2+
\eta_{RR}^2)u^2+\frac{1}{4}(\eta_{LR}^2+\eta_{RL}^2)(s^2+t^2),
\end{equation}
where $s$, $t$ and $u$ are the usual Lorentz-invariant variables. In the 
$s$ channel one has $t=-s(1-\cos\theta)/2$, $u=-s(1+\cos\theta)/2$.
For definiteness,
here and later on as the scattering angle $\theta$ there will be chosen
that  between the second particles in the initial and final pairs. The cross
section for the $u$ channel corresponds to substitution $s\leftrightarrow u$
on the right hand side of Eq.~(\ref{section1}).

The main background to these four-lepton LFV processes is given by the
family conserving  six-lepton SM processes with the same charged lepton content
and with the imbalance of family compensated by a pair of corresponding 
neutrinos. More particularly, for the processes of Eq.~(\ref{l'lll-processes}) 
they are
\begin{eqnarray}
e^+e^-&\to&\mu^+e^-(\nu_\mu\bar\nu_e),\;
e^+\mu^-(\bar\nu_\mu\nu_e),\nonumber\\
\mu^+e^-&\to&e^+e^-(\bar\nu_\mu\nu_e),\nonumber\\
e^-e^-&\to&\mu^-e^-(\bar\nu_\mu\nu_e),\\
\mu^-e^-&\to&e^-e^-(\nu_\mu\bar\nu_e).\nonumber
\end{eqnarray}
We expect this background to be rather small due to the back-to-back 
requirement
for charged final leptons, as well as the requirement for missing energy
$\not\!\!E<2\Delta E$, where $\Delta E$ is the energy uncertainty 
of charged final leptons.
Nevertheless this background should be evaluated more thoroughly. Not knowing
the characteristics of the detectors one can hardly do it. 
Hence the question arises as 
to whether this background, if present and dangerous, could be suppressed by 
using the polarized beams. The answer is that it can really be done so to a
large extent. Let us consider this topic in more detail.

The typical Feynman diagrams for these background processes are given in 
Fig.~1. As it is seen, there are two topologically different classes of 
diagrams: those
with and without the triple-boson vertex, one of the bosons in the vertex
being neutral with necessity. On the other
hand, the second class is divided into two subclasses: those with and without 
neutral
boson exchange. The same classification takes place for the six-lepton SM
background, if allowed, for all other LFV processes to be considered. It is
clear that the neutral boson diagrams could be encountered only in processes 
with a 
particle and antiparticle of the same type, or with a going-through particle,
as it is indeed the case with the $l'lll$ type of the LFV processes.
These observations are important in discussing the possible  background 
reduction by the initial state polarizations. 

Suppose that arbitrary longitudinal polarization $P$
can be achieved. In this case, by proper choosing the ``unnatural'' polarization 
of one of the initial particles ({\it i.e.}, $P>0$ for leptons, or $P<0$ for 
antileptons) one can suppress as $\Delta P\equiv1-|P|$ all the background 
diagrams not containing the neutral 
bosons\footnote{The problem with rotating the ``natural''  polarization
of the produced muons to the required unnatural one can be hoped to be 
eventually solved.}. What about the diagrams with the latter boson,
they can be suppressed only if this 
boson is not coupled to the initial particles. But there are always diagrams 
with such a coupling and hence the background cannot be completely reduced 
by using just one polarized beam. Fortunately, by using the second polarized 
beam, again with the unnatural polarization, one can suppress all the left-out
diagrams as well, but for the diagrams  with the triple-boson vertex in 
the $s$ channel annihilation. 
The degree of the double polarization suppression will
be just $\Delta P$ (instead of $\Delta P^2$ what might be expected
{\it a priori}). Nevertheless the six-lepton SM background, if allowed, 
can rather efficiently be reduced in the bulk of the processes. 
More than this, there are  ``clean'' processes without the six-lepton 
SM background at all (see later on)\footnote{We consider
as even less probable the six-lepton SM background with some of the charged 
leptons missed as,
say, in $e^+e^-\to \mu^+e^-(\not\!\!e^+\!\!\not\!\!\!\mu^-)$, etc, or 
the eight-fermion
SM background with missing quarks as, say, $e^+e^-\to s\bar s \to \mu^+e^- 
(\nu_\mu\bar\nu_e\not\!u\not\!\bar u)$, etc.}.

The $\mu eee$ class of the LFV processes is also unfavourable in another aspect. 
Indeed, one can show that
\begin{equation}
B(\mu\to ee\bar e)=\frac{1}{4}\biggl(\frac{v}{F}\biggr)^4\biggl(\eta_{LL}^2+
\eta_{RR}^2+\frac{1}{2}(\eta_{LR}^2+\eta_{RL}^2)\biggr),
\end{equation}
with $v\equiv(\sqrt{2}G_F)^{-1/2}$ being the v.e.v. and  
$G_F$ being the Fermi constant. From\footnote{All the available 
restrictions on the LFV decay branchings are in what follows  
from Ref.~\cite{rpp} and correspond to 90\%~C.L.} 
$B(\mu\to ee\bar e)< 10^{-12}$ it follows that $F> 2^{3/4}\,(\eta_{LL}^2+
\eta_{RR}^2+(\eta_{LR}^2+\eta_{RL}^2)/2)^{1/4}\, 10^2$\,TeV.
So, one expects typically for the LFV processes of Eq.~(\ref{l'lll-processes}) 
that
$\sigma<10^{-3}$\,fb, and hence at $L=10^3$\,fb$^{-1}$ one gets no more than
1~event/year. Hence this case is marginally observable and constitutes in a
sense the worst case study. For the rest of the four-lepton LFV
processes there are either no experimental restrictions at all or
just rather mild ones, $F> {\cal O}(1$\,TeV). Hence all of these processes can 
in principle be reliably observed. 

Before going to the next type of the LFV processes let us 
complete the discussion of the left-out classes of the $l'lll$ type.

\boldmath
{\bf ({\bf\em ii}) $e\mu\mu\mu$}\unboldmath\ $\;$
This class could contribute to the following LFV pro\-ces\-ses
\begin{eqnarray}
&(s)&\;\;\;\;\;\cases{\mu^+\mu^-\to e^+\mu^-,\;\mu^+e^-,\cr
		    \mu^+e^-\to \mu^+\mu^-,\cr}\nonumber\\
&(u)&\;\;\;\;\;\cases{\mu^-\mu^-\to e^-\mu^-,\cr
                     e^-\mu^-\to \mu^-\mu^-.\cr}
\end{eqnarray}
All the preceding discussion of the $\mu eee$ class is unchanged but for the 
absence of experimental restrictions on the scale $F$. Needless to say that
here the coefficients $\eta$ in the counterpart of Eq.~(\ref{section1})
are some new free parameters for each of the classes within the type. The same 
goes without saying in what follows.

\boldmath
{\bf ({\bf\em iii}) $\tau eee$}\unboldmath\ $\;$
The relevant LFV processes are
\begin{eqnarray}
&(s)&\;\;\;\;\;e^+e^-\to\tau^+e^-,\;e^+\tau^-,\\
&(u)&\;\;\;\;\;e^-e^-\to\tau^-e^-.\nonumber
\end{eqnarray}
The cross sections and background are similar to the $\mu eee$ case. From
the limit $B(\tau\to ee\bar e)<3\cdot10^{-6}$ it follows the mild restriction
$F>{\cal O}(1$\,TeV).

\boldmath
{\bf ({\bf\em iv}) $\tau\mu\mu\mu$}\unboldmath\ $\;$
The LFV processes are
\begin{eqnarray}
&(s)&\;\;\;\;\;\mu^+\mu^-\to\tau^+\mu^-,\;\mu^+\tau^-,\\
&(u)&\;\;\;\;\;\mu^-\mu^-\to\tau^-\mu^-,\nonumber
\end{eqnarray}
and from $B(\tau\to\mu\mu\bar\mu)<1.9\cdot10^{-6}$ it follows the same 
restriction on  $F$. Cross sections and background are as before.
Altogether, these four classes of the LFV processes depend on 16 
free parameters.

\paragraph{(2) \boldmath $l'l'll$ type \unboldmath} 

The most general LFV effective lagrangian of this type is given by\footnote{
Operators of the form $(\bar l'l')(\bar ll)$ do not violate family and are 
omitted.} 
\begin{equation}\label{l'l'll}
{\cal L}=\frac{1}{2F^2}\biggl(c_{LL}(\bar l'l)_L\,(\bar l'l)_L
+c_{LR}(\bar l'l)_L\,(\bar l'l)_R+(L\leftrightarrow R)+\mbox{h.c.}\biggr)
\end{equation}
with $c_{LR}=c_{RL}$. Here 1/2 in front of the expression is introduced to 
account for the
identity of currents. One encounters three particular classes: $\mu\mu ee$,
$\tau\tau ee$ and $\tau\tau\mu\mu$.

\boldmath
{\bf ({\bf\em i}) $\mu\mu ee$}\unboldmath\ $\;$
Here one has $l'=\mu$ and $l=e$. These interactions contribute to the following
LFV reactions
\begin{eqnarray}
&(s)&\;\;\;\;\;\;\;\;\mu^+e^-\to e^+\mu^-,\nonumber\\
&(u)&\;\;\;\;\;\cases{e^-e^-\to \mu^-\mu^-,\cr
                    \mu^-\mu^-\to e^-e^-.\cr}
\end{eqnarray}
They are intimately related with the muonium-antimuonium ($M\bar M$) 
conversion 
$\mu^+e^-\leftrightarrow e^+\mu^-$~\cite{ponte,fein} and were discussed 
in the context of the 
multiplicative conservation laws in Refs.~\cite{glash,hou}\footnote{The
effective lagrangian of Eq.~(\ref{l'l'll}) is the most general one 
consistent with
the SM invariance, and as such it can be used for the model independent analysis
of the $M\bar M$ conversion.}.

The $s$ channel cross section is
\begin{equation}
8\pi s F^4\frac{d\sigma}{d\cos\theta}=(c_{LL}^2+
c_{RR}^2)u^2+\frac{1}{2}c_{LR}^2(s^2+t^2),
\end{equation}
and that for the $u$ channel corresponds to the substitution 
$s\leftrightarrow u$ on the right hand side.

There is no six-lepton SM background for all these processes, as well as for 
all others of the same type. This is because $|\Delta F_{l,l'}|=2$ for two 
different 
types of leptons $l,l'$ cannot be compensated simultaneously just by emission 
of two
additional neutrinos, at least four of them being required. So the leading 
SM background here is given by the eight-lepton processes
\begin{eqnarray}
\mu^+e^-&\to&e^+\mu^-(\nu_e\nu_e\bar\nu_\mu\bar\nu_\mu),
\nonumber\\
e^-e^-&\to&\mu^-\mu^-(\nu_e\nu_e\bar\nu_\mu\bar\nu_\mu),\\
\mu^-\mu^-&\to&e^-e^-(\bar\nu_e\bar\nu_e\nu_\mu\nu_\mu).\nonumber
\end{eqnarray}
It can be shown that among the corresponding Feynman diagrams
there always exist those with emission of the neutral gauge boson by the
initial particles. Hence the eight-lepton background cannot be completely
suppressed by using just one polarized beam, double polarization being 
required. But we expect this background
not to be fatal because of the smallness of its cross section, as well as
the difficulty to concentrate almost all the energy just in two of the
six final particles.

What about the experimental restriction on this class of interactions, it 
can be inferred only from the limit $G_{M\bar M}<1.8\cdot10^{-2}
G_F$~\cite{mmbar} on the four-fermion effective coupling for the $M\bar M$ 
conversion  and corresponds
to $F> {\cal O}(1$\,TeV). This still leaves a lot of room to investigate at 
the future high energy lepton colliders, the $\mu^-\mu^-/\mu^+\mu^+$
collider with required advanced parameters being  be most realistic
among three conceivable ones.

\boldmath
{\bf ({\bf\em ii}) $\tau\tau ee$}\unboldmath\ $\;$
The only relevant LFV process in this class is
\begin{eqnarray}
e^-e^-&\to&\tau^-\tau^-.
\end{eqnarray}
There is neither six-lepton SM background nor experimental restrictions on $F$.

\boldmath
{\bf ({\bf\em iii}) $\tau\tau\mu\mu$}\unboldmath\ $\;$
The similar LFV process here is
\begin{eqnarray}
\mu^-\mu^-&\to&\tau^-\tau^-.
\end{eqnarray}
Everything is the same as in the preceding case\footnote{Note that the last two
processes are supplementary to those of the $\mu\mu ee$ class w.r.t.\ the study
of the multiplicative conservation laws. Were these laws exact, the processes 
of the $l'l'll$ type could nevertheless be always allowed independent of the 
particular 
assignment of the multiplicative quantum numbers. On the other hand, the
allowance of the various interactions of the $l'lll$ and $l''l'll$ types
depends crucially on the latter choice. This makes it obligatory to investigate 
experimentally all the conceivable channels of the LFV processes.}.
This type of the LFV interactions depends altogether on 9 free parameters.

\paragraph{(3) \boldmath $l''l'll$ type\unboldmath}

This is the most general type of the four-lepton LFV interactions. The general
effective lagrangian here is\footnote{Operators of the form $(\bar l''l)_L
(\bar ll')_L$ are Fiertz-equivalent to those $(\bar l''l')_L
(\bar ll)_L$, and $(\bar l''l)_R(\bar ll')_R$ to $(\bar l''l')_R(\bar ll)_R$.}
\begin{eqnarray}\label{l'l''ll}
{\cal L}&=&\frac{1}{F^2}\biggl(c_{LL}^{(1)}(\bar l''l)_L\,(\bar l'l)_L
+ c_{LR}^{(1)}(\bar l''l)_L\,(\bar l'l)_R\nonumber\\
&&\;\;\;+\; c_{LL}^{(2)}(\bar l''l')_L\,(\bar ll)_L
+ c_{LR}^{(2)}(\bar l''l')_L\,(\bar ll)_R\\
&&\;\;\;+\; c_{LR}^{(3)}(\bar l''l)_L\,(\bar ll')_R +
(L\leftrightarrow R)+\mbox{h.c.}\biggr)\nonumber
\end{eqnarray}
Three different classes are conceivable: $\tau\mu ee$, $\tau e \mu\mu$
and $e\mu\tau\tau$.

\boldmath
{\bf ({\bf\em i}) $\tau\mu ee$}\unboldmath\ $\;$
Here one has $l''=\tau$, $l'=\mu$, $l=e$, and the following LFV processes are
allowed
\begin{eqnarray}\label{l''l'llprocesses1}
&(s)&\;\;\;\;\;e^+e^-\to\tau^+\mu^-,\;\mu^+\tau^-,\nonumber\\
&(t)&\;\;\;\;\;\mu^+e^-\to\tau^+e^-,\\
&(u)&\;\;\;\;\;\mu^-e^-\to e^-\tau^-,\nonumber
\end{eqnarray}
as well as
\begin{eqnarray}\label{l''l'llprocesses2}
&(s)&\;\;\;\;\;\mu^+e^-\to e^+\tau^-,\nonumber\\
&(u)&\;\;\;\;\;e^-e^-\to\mu^-\tau^-.
\end{eqnarray}

The $s$ channel cross section for the  processes of 
Eq.~(\ref{l''l'llprocesses1}) is
\begin{eqnarray}\label{cross-section1}
32\pi s F^4\frac{d\sigma}{d\cos\theta}&=&
(c_{LL}^{(2)\,2}+c_{RR}^{(2)\,2})u^2\nonumber\\
&+&(c_{LR}^{(2)\,2}+c_{RL}^{(2)\,2})t^2
+(c_{LR}^{(3)\,2}+c_{RL}^{(3)\,2})s^2,
\end{eqnarray}
the other two being obtained by substitutions, respectively, 
$s\leftrightarrow t$ and $s\leftrightarrow u$ on the right hand side. 
The cross section for the first process of
Eq.~(\ref{l''l'llprocesses2}) is
\begin{equation}\label{cross-section2}
8\pi s F^4 \frac{d\sigma}{d\cos\theta}=
(c_{LL}^{(1)\,2}+c_{RR}^{(1)\,2})u^2
+\frac{1}{4}(c_{LR}^{(1)\,2}+c_{RL}^{(1)\,2})(s^2+t^2),
\end{equation}
with that for the second process being obtained by  substitution 
$s\leftrightarrow u$ on the right hand side.

The six-lepton SM background is allowed for the processes of 
Eq.~(\ref{l''l'llprocesses1}) but is forbidden for those of 
Eq.~(\ref{l''l'llprocesses2}). The latter fact is because it is impossible to 
compensate $|\Delta F_l|=2$ for leptons of the kind $l$, plus additional
LFV $|\Delta F_{l',l''}|=1$ for $l'$ and $l''$, just by emission of two 
neutrinos. Nevertheless the eight-lepton SM background with family 
conservation is as always
allowed. Finally, from $B(\tau\to\mu e\bar e,\bar\mu ee)<6.8\cdot10^{-6}$ 
it follows that $F>{\cal O}(1\,$TeV).

\boldmath
{\bf ({\bf\em ii}) $\tau e\mu\mu$}\unboldmath\ $\;$
It corresponds to  $l''=\tau$, $l'=e$, $l=\mu$, and the relevant LFV 
processes here are
\begin{eqnarray}
&(s)&\;\;\;\;\;\mu^+\mu^-\to\tau^+e^-,\;e^+\tau^-,\nonumber\\
&(t)&\;\;\;\;\;\mu^+e^-\to\mu^+\tau^-,\\
&(u)&\;\;\;\;\;e^-\mu^-\to\mu^-\tau^-\nonumber
\end{eqnarray}
and
\begin{eqnarray}
&(s)&\;\;\;\;\;\mu^+e^-\to \tau^+\mu^-,\nonumber\\
&(u)&\;\;\;\;\;\mu^-\mu^-\to e^-\tau^-.
\end{eqnarray}
The cross sections and backgrounds are similar to the preceding case. From 
$B(\tau\to e\mu\bar\mu,\bar e\mu\mu)<7.1\cdot10^{-6}$ it follows the same 
lower bound $F>{\cal O}(1$\,TeV).

\boldmath
{\bf ({\bf\em iii}) $e\mu\tau\tau$}\unboldmath\ $\;$
Here one has $l''=e$, $l'=\mu$ and $l=\tau$. The processes of interest are
\begin{eqnarray}\label{ttme1}
&(s)&\;\;\;\;\;\mu^+e^-\to \tau^+\tau^-
\end{eqnarray}
 and
\begin{eqnarray}\label{ttme2}
&(u)&\;\;\;\;\;\mu^-e^-\to \tau^-\tau^-.
\end{eqnarray}
Eqs.~(\ref{cross-section1}) and (\ref{cross-section2}) with the proper choice 
for invariant variables are relevant here, too. The six-lepton SM background 
to the process of Eq.~(\ref{ttme1}) is as before. That one for the 
process of Eq.~(\ref{ttme2}) is again absent, only the eight-lepton background
being allowed.
Besides, there are no experimental restrictions on this class of 
interactions\footnote{We considered everywhere only the SM background. 
It is beyond our scope to discuss the possible instrumental background due to  
faking  one kind of particles by another, or caused, say, by cosmic rays, etc. 
Because of the extreme rarity of the LFV events every conceivable
background should be taken into account eventually.
Hence what we have found are the most optimistic estimates.}. 
This last type of the LFV processes contains a total of 30 free parameters.

\paragraph{Discussion and conclusion}

Our general classification of the four-lepton LFV processes is summarized in
Table~1.
In Table~2 we present the helicity differential cross sections for the 
four-lepton LFV processes in the units of $(8\pi sF^4)^{-1}$ 
(the upper expressions in 
each of the cellars). Let us remind that the coefficients 
$\eta$, $c$ and $c^{(i)}$ for the $l'lll$, $l'l'll$ and $l''l'll$ types of the
LFV processes, respectively, differ for various classes of the processes.
Nevertheless they are designated in the same manner for simplicity. 
As it should be clear from the preceding exposition, 
all four-lepton LFV processes depend in general on the grand total of 
55 free parameters.

One can easily infer from the Table~2 which particular four-lepton LFV 
operators could be studied in each of the channels. In the SM background 
clean channels
all the contributing interactions could be studied equally well. In the 
channels with the six-lepton SM background the interactions contributing 
to the helicity channels with the unnatural initial polarizations could be
studied more reliably.

The lower expressions in each of the cellars of the Table~2 represent  the 
reference number of the LFV events 
$ n_{\lambda_1\lambda_2}$ in helicity
channels at the proper collider with $\sqrt{s}=4$\,TeV and $L=10^3$\,fb$^{-1}$
for the flavour/family scale $F=10^2$\,TeV.
The real number of the events $N_{\lambda_1\lambda_2}$ 
scales  with $F$, $\sqrt{s}$ and $L$ as $(10^2 $TeV$/F)^4
(\sqrt{s}/4\,$TeV)$^2\,(L/10^3\,$fb$^{-1})\,n_{\lambda_1\lambda_2}$. 
For the unpolarized beams one has
$N=\sum  N_{\lambda_1\lambda_2}/4$. If one takes as an observability 
criterion that the number of LFV events in a channel is equal, say, to 10 then
the LFV mass scale that can be reached in the channel is 
$F=(\sqrt{s}/4\,\mbox{TeV})^{1/2}\,(L/10^3\,\mbox{fb}^{-1})^{1/4}\,
( n_{\lambda_1\lambda_2}/10)^{1/4}\,10^2$\,TeV.

We conclude that the LFV processes are quite amenable to experimental study 
at the future generation lepton colliders 
in the most favourable cases up to the underlying  scale ${\cal O}(10^2$\,TeV).
Studying these processes 
to solve the outstanding flavour/family problem might provide 
{\it raison d'\^{e}tre} and the great  challenge for the colliders.

\paragraph{Acknowledgements}
This work is supported in part by the RFBR under grant No.~96-02-18122 and in
part by the Competition Centre for Fundamental Natural Sciences under 
grant No.~95-0-6.4-21.     
For drawing the Feynman diagrams use is made of the D\"{U}RER
graphical package~\cite{durer}.

%\newpage
%\section*{Figure caption}
%\paragraph{Fig.~1} The typical Feynman diagrams of the  family conserving
%six-lepton SM background for the four-lepton LFV processes of the $\mu eee$ 
%class.

\newpage
\begin{table}[h]
\begin{tabular}{|c|c|l|c|c|}
\hline 
Type&Class&Processes&Restriction&Background\\
\hline
&&$e^+e^-\to\mu^+e^-,\;e^+\mu^-$&&\\
&&$\mu^+e^-\to e^+e^-$&&\\
&$\mu eee$&$e^-e^-\to\mu^-e^-$&${\cal O}(10^2$\,TeV)&\\
&&$\mu^-e^-\to e^-e^-$&&\\
\cline{2-4}
&&$\mu^+\mu^-\to e^+\mu^-,\;\mu^+e^-$&&\\
&&$\mu^+e^-\to \mu^+\mu^-$&&\\
$l'lll$&$e\mu\mu\mu$&$\mu^-\mu^-\to\mu^-e^-$&No&Yes\\
&&$e^-\mu^-\to\mu^-\mu^-$&&\\
\cline{2-4}
&&$e^+e^-\to\tau^+e^-,\;e^+\tau^-$&&\\
&$\tau eee$&$e^-e^-\to\tau^-e^-$&${\cal O}(1$\,TeV)&\\
\cline{2-4}
&&$\mu^+\mu^-\to\tau^+\mu^-,\;\mu^+\tau^-$&&\\
&$\tau\mu\mu\mu$&$\mu^-\mu^-\to\tau^-\mu^-$&${\cal O}(1$\,TeV)&\\
\hline
&&$\mu^+e^-\to e^+\mu^-$&&\\
&$\mu\mu ee$&$e^-e^-\to \mu^-\mu^-$&${\cal O}(1$\,TeV)&\\
$l'l'll$&&$\mu^-\mu^-\to e^-e^-$&&No\\
\cline{2-4}
&$\tau\tau ee$&$e^-e^-\to\tau^-\tau^-$&No&\\
\cline{2-4}
&$\tau\tau\mu\mu$&$\mu^-\mu^-\to\tau^-\tau^-$&No&\\
\hline
&&$e^+e^-\to\tau^+\mu^-,\;\mu^+\tau^-$&&\\
&&$\mu^+e^-\to\tau^+e^-$&${\cal O}(1$\,TeV)&Yes\\
&$\tau\mu ee$&$\mu^-e^-\to e^-\tau^-$&&\\
\cline{3-5}
&&$\mu^+e^-\to e^+\tau^-$&&\\
&&$e^-e^-\to\mu^-\tau^-$&${\cal O}(1$\,TeV)&No\\
\cline{2-5}
&&$\mu^+\mu^-\to\tau^+e^-,\;e^+\tau^-$&&\\
$l''l'll$&&$\mu^+e^-\to\mu^+\tau^-$&${\cal O}(1$\,TeV)&Yes\\
&$\tau e\mu\mu$&$e^-\mu^-\to\mu^-\tau^-$&&\\
\cline{3-5}
&&$\mu^+e^-\to \tau^+\mu^-$&&\\
&&$\mu^-\mu^-\to e^-\tau^-$&${\cal O}(1$\,TeV)&No\\
\cline{2-5}
&$\mu e\tau\tau$&$\mu^+e^-\to \tau^+\tau^-$&No&Yes\\
\cline{3-5}
&&$\mu^-e^-\to \tau^-\tau^-$&No&No\\
\hline
\end{tabular}
\paragraph{Table~1} Summary of the four-lepton LFV processes (see text).
\end{table}

\newpage
\begin{table}[h]
\begin{tabular}{|c|l|c|c|c|c|}
\hline
&&\multicolumn{4}{c}{$8\pi sF^4d\sigma_{\lambda_1\lambda_2}
/d\cos\theta$}\vline\\
Type&Processes&\multicolumn{4}{c}{$ n_{\lambda_1\lambda_2}$}\vline\\
\cline{3-6}
&&$-\;-$&$+\;+$&$-\;+$&$+\;-$\\ 
\hline 
&$e^+e^-\to \mu^+e^-/\tau^+e^-$&$\eta_{RL}^2s^2$&$\eta_{LR}^2s^2$&
$4\eta_{RR}^2u^2+\eta_{LR}^2t^2$ &$4\eta_{LL}^2u^2+\eta_{RL}^2t^2$ \\ [2mm]
&$\mu^+\mu^-\to e^+\mu^-/\tau^+\mu^-$&$5\eta_{RL}^2$&$5\eta_{LR}^2$&
$1.7(4\eta_{RR}^2+\eta_{LR}^2)$&$1.7(4\eta_{LL}^2+\eta_{RL}^2)$\\ [3mm]
\cline{2-6}
&$e^+e^-\to e^+\mu^-/e^+\tau^-$&$\eta_{LR}^2s^2$&$\eta_{RL}^2s^2$&
$4\eta_{RR}^2u^2+\eta_{LR}^2t^2$ &$4\eta_{LL}^2u^2+\eta_{RL}^2t^2$ \\ [2mm]
&$\mu^+\mu^-\to \mu^+e^-/\mu^+\tau^-$&$5\eta_{LR}^2$&$5\eta_{RL}^2$&
$1.7(4\eta_{RR}^2+\eta_{LR}^2)$&$1.7(4\eta_{LL}^2+\eta_{RL}^2)$\\ [3mm]
\cline{2-6}
&$\mu^+e^-\to e^+e^-$&$\eta_{RL}^2s^2$&$\eta_{LR}^2s^2$&
$4\eta_{RR}^2u^2+\eta_{RL}^2t^2$ *&$4\eta_{LL}^2u^2+\eta_{LR}^2t^2$\\ [2mm]
&&$5\eta_{RL}^2$&$5\eta_{LR}^2$&$1.7(4\eta_{RR}^2+\eta_{RL}^2)$&
$1.7(4\eta_{LL}^2+\eta_{LR}^2)$\\ [3mm]
\cline{2-6}
$l'lll$&$\mu^+e^-\to \mu^+\mu^-$&$\eta_{LR}^2s^2$&$\eta_{RL}^2s^2$&
$4\eta_{RR}^2u^2+\eta_{RL}^2t^2$ *&$4\eta_{LL}^2u^2+\eta_{LR}^2t^2$\\[2mm]
&&$5\eta_{LR}^2$&$5\eta_{RL}^2$&$1.7(4\eta_{RR}^2+\eta_{RL}^2)$&
$1.7(4\eta_{LL}^2+\eta_{LR}^2)$\\ [3mm]
\cline{2-6}
&$e^-e^-\to\mu^-e^-/\tau^-e^-$&$4\eta_{LL}^2s^2$&$4\eta_{RR}^2s^2$ *&
$\eta_{LR}^2u^2+\eta_{RL}^2t^2$&$\eta_{LR}^2t^2+\eta_{RL}^2u^2$\\ [2mm]
&$\mu^-\mu^-\to e^-\mu^-/\tau^-\mu^-$&$20\eta_{LL}^2$&$20\eta_{RR}^2$&
$1.7(\eta_{LR}^2+\eta_{RL}^2)$&$1.7(\eta_{LR}^2+\eta_{RL}^2)$\\ [3mm]
\cline{2-6}
&$\mu^-e^-\to e^-e^-$&$4\eta_{LL}^2s^2$&$4\eta_{RR}^2s^2$ *&
$\eta_{LR}^2(u^2+t^2)$&$\eta_{RL}^2(u^2+t^2)$\\ [2mm]
&$e^-\mu^-\to \mu^-\mu^-$&$10\eta_{LL}^2$&$10\eta_{RR}^2$&
$1.7\eta_{LR}^2$&$1.7\eta_{RL}^2$\\ [3mm]
\hline
&$e^+e^-\to\tau^+\mu^-$&$c_{RL}^{(3)\,2}s^2$&$c_{LR}^{(3)\,2}s^2$&
$c_{RR}^{(2)\,2}u^2+c_{LR}^{(2)\,2}t^2$ &
$c_{LL}^{(2)\,2}u^2+c_{RL}^{(2)\,2}t^2$\\ [2mm]
&$\mu^+\mu^-\to\tau^+e^-$&$5c_{RL}^{(3)\,2}$&$5c_{LR}^{(3)\,2}$&
$1.7(c_{RR}^{(2)\,2}+c_{LR}^{(2)\,2})$&
$1.7(c_{LL}^{(2)\,2}+c_{RL}^{(2)\,2})$\\ [3mm]
\cline{2-6}
&$e^+e^-\to\mu^+\tau^-$&$c_{LR}^{(3)\,2}s^2$&$c_{RL}^{(3)\,2}s^2$&
$c_{RR}^{(2)\,2}u^2+c_{LR}^{(2)\,2}t^2$ &
$c_{LL}^{(2)\,2}u^2+c_{RL}^{(2)\,2}t^2$\\ [2mm]
&$\mu^+\mu^-\to e^+\tau^-$&$5c_{LR}^{(3)\,2}$&$5c_{RL}^{(3)\,2}$&
$1.7(c_{RR}^{(2)\,2}+c_{LR}^{(2)\,2})$&
$1.7(c_{LL}^{(2)\,2}+c_{RL}^{(2)\,2})$\\ [3mm]
\cline{2-6}
&$\mu^+e^-\to\tau^+e^-$&$c_{RL}^{(2)\,2}s^2$&$c_{LR}^{(2)\,2}s^2$&
$c_{RR}^{(2)\,2}u^2+c_{LR}^{(3)\,2}t^2$ *&
$c_{LL}^{(2)\,2}u^2+c_{RL}^{(3)\,2}t^2$\\ [2mm]
&&$5c_{RL}^{(2)\,2}$&$5c_{LR}^{(2)\,2}$&
$1.7(c_{RR}^{(2)\,2}+c_{LR}^{(3)\,2})$&
$1.7(c_{LL}^{(2)\,2}+c_{RL}^{(3)\,2})$\\ [3mm]
\cline{2-6}
$l''l'll$&$\mu^+e^-\to\mu^+\tau^-$&$c_{LR}^{(2)\,2}s^2$&$c_{RL}^{(2)\,2}s^2$&
$c_{RR}^{(2)\,2}u^2+c_{LR}^{(3)\,2}t^2$ *&
$c_{LL}^{(2)\,2}u^2+c_{RL}^{(3)\,2}t^2$\\ [2mm]
&&$5c_{LR}^{(2)\,2}$&$5c_{RL}^{(2)\,2}$&
$1.7(c_{RR}^{(2)\,2}+c_{LR}^{(3)\,2})$&
$1.7(c_{LL}^{(2)\,2}+c_{RL}^{(3)\,2})$\\ [3mm]
\cline{2-6}
&$\mu^+e^-\to\tau^+\tau^-$&$c_{RL}^{(3)\,2}s^2$&$c_{LR}^{(3)\,2}s^2$&
$c_{RR}^{(2)\,2}u^2+c_{RL}^{(2)\,2}t^2$ *&
$c_{LL}^{(2)\,2}u^2+c_{LR}^{(2)\,2}t^2$\\ [2mm]
&&$5c_{RL}^{(3)\,2}$&$5c_{LR}^{(3)\,2}$&
$1.7(c_{RR}^{(2)\,2}+c_{RL}^{(2)\,2})$&
$1.7(c_{LL}^{(2)\,2}+c_{LR}^{(2)\,2})$\\ [3mm]
\cline{2-6}
&$\mu^-e^-\to e^-\tau^-$&$c_{LL}^{(2)\,2}s^2$&$c_{RR}^{(2)\,2}s^2$ *&
$c_{RL}^{(3)\,2}u^2+c_{RL}^{(2)\,2}t^2$&
$c_{LR}^{(3)\,2}u^2+c_{LR}^{(2)\,2}t^2$\\ [2mm]
&$e^-\mu^-\to\mu^-\tau^-$&$5c_{LL}^{(2)\,2}$&$5c_{RR}^{(2)\,2}$&
$1.7(c_{RL}^{(3)\,2}+c_{RL}^{(2)\,2})$&
$1.7(c_{LR}^{(3)\,2}+c_{LR}^{(2)\,2})$\\ [3mm]
\hline
\end{tabular}
\paragraph{Table 2}Helicity differential cross sections for the LFV processes 
in the units of $(8\pi sF^4)^{-1}$ (upper entries) and the representative 
number of events $n_{\lambda_1\lambda_2}$ in the various helicity
channels at $F=10^2$\,TeV, $\sqrt{s}=4$\,TeV and $L=10^3$\,fb$^{-1}$ (lower
entries). {\bf (a)} The processes with the six-lepton  SM background.
The asterisk  $\ast$ marks the unnatural initial helicity channels  
without six-lepton  SM background: the $+\,+$ helicity lepton-lepton collisions
and the $-\,+$ helicity antilepton-lepton collisions (see text).
\end{table}

\begin{table}[h]
\begin{tabular}{|c|l|c|c|c|c|}
\hline
&&\multicolumn{4}{c}{$8\pi sF^4d\sigma_{\lambda_1\lambda_2}
/d\cos\theta$}\vline\\
Type&Processes&\multicolumn{4}{c}{$ n_{\lambda_1\lambda_2}$}\vline\\
\cline{3-6}
&&$-\;-$&$+\;+$&$-\;+$&$+\;-$\\
\hline
&$\mu^+e^-\to e^+\mu^-$ &$c_{LR}^2s^2$&$c_{LR}^2s^2$&
$4c_{RR}^2u^2+c_{LR}^2t^2$&$4c_{LL}^2u^2+c_{LR}^2t^2$\\ [2mm]
&&$5c_{LR}^2$&$5c_{LR}^2$&$1.7(4c_{RR}^2+c_{LR}^2)$&
$1.7(4c_{LL}^2+c_{LR}^2)$\\ [3mm]
\cline{2-6}
$l'l'll$&$e^-e^-\to \mu^-\mu^-/\tau^-\tau^-$ &$4c_{LL}^2s^2$&$4c_{RR}^2s^2$&
$c_{LR}^2(u^2+t^2)$&$c_{LR}^2(u^2+t^2)$\\ [2mm]
&$\mu^-\mu^-\to e^-e^-/\tau^-\tau^-$ &$10c_{LL}^2$&$10c_{RR}^2$&
$1.7c_{LR}^2$&$1.7c_{LR}^2$\\ [3mm]
\hline
&$\mu^+e^-\to e^+\tau^-$ &$c_{LR}^{(1)\,2}s^2$&
$c_{RL}^{(1)\,2}s^2$&$4c_{RR}^{(1)\,2}u^2
+c_{LR}^{(1)\,2}t^2$&$4c_{LL}^{(1)\,2}u^2
+c_{RL}^{(1)\,2}t^2$\\ [2mm]
&$\mu^+e^-\to\tau^+\mu^-$ &$5c_{LR}^{(1)\,2}$&$5c_{RL}^{(1)\,2}$&
$1.7(4c_{RR}^{(1)\,2}+c_{LR}^{(1)\,2})$&
$1.7(4c_{LL}^{(1)\,2}+c_{RL}^{(1)\,2})$\\ [3mm]
\cline{2-6}
$l''l'll$&$e^-e^-\to \mu^-\tau^-$ &$4c_{LL}^{(1)\,2}s^2$&$4c_{RR}^{(1)\,2}s^2$&
$c_{LR}^{(1)\,2}t^2+c_{RL}^{(1)\,2}u^2$&
$c_{LR}^{(1)\,2}u^2+c_{RL}^{(1)\,2}t^2$\\ [2mm]
&$\mu^-\mu^-\to\tau^-e^-$ &$20c_{LL}^{(1)\,2}$&$20c_{RR}^{(1)\,2}$&
$1.7(c_{LR}^{(1)\,2}+c_{RL}^{(1)\,2})$&
$1.7(c_{LR}^{(1)\,2}+c_{RL}^{(1)\,2})$\\ [3mm]
\cline{2-6}
&$\mu^-e^-\to \tau^-\tau^-$ &$4c_{LL}^{(1)\,2}s^2$&$4c_{RR}^{(1)\,2}s^2$&
$c_{LR}^{(1)\,2}(t^2+u^2)$&$c_{RL}^{(1)\,2}(t^2+u^2)$\\ [2mm]
&&$10c_{LL}^{(1)\,2}$&$10c_{RR}^{(1)\,2}$&
$1.7c_{LR}^{(1)\,2}$&$1.7c_{RL}^{(1)\,2}$\\ [3mm]
\hline
\end{tabular}
\paragraph{Table 2} Continued. ({\bf b)} The  processes clean of the 
six-lepton SM background  (see text).
\end{table}

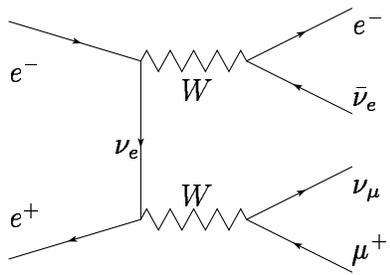
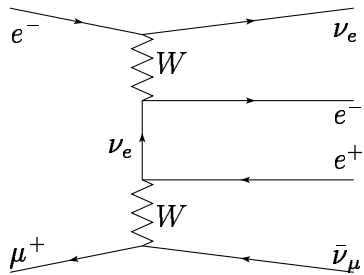
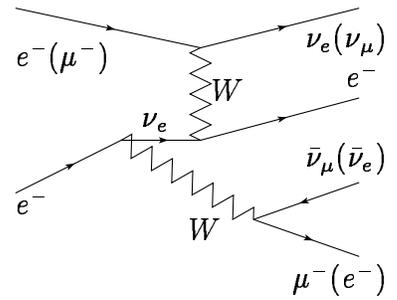
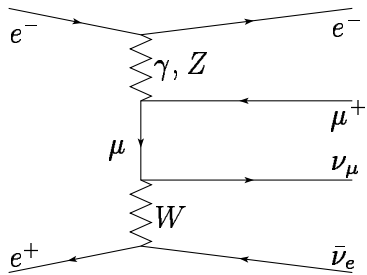
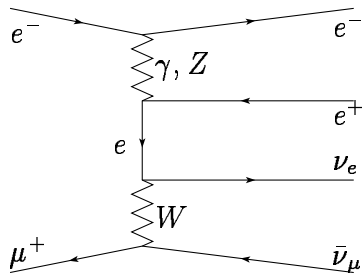
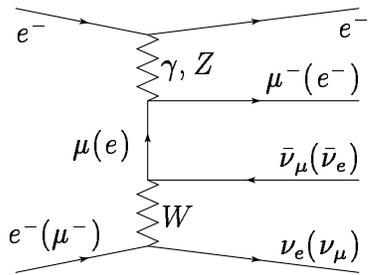
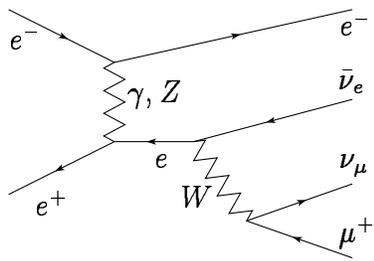
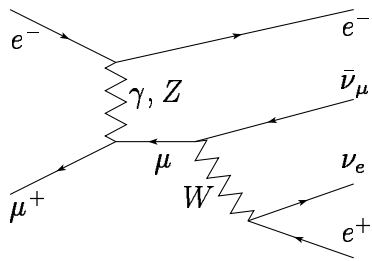
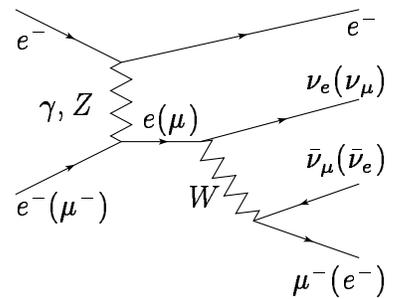
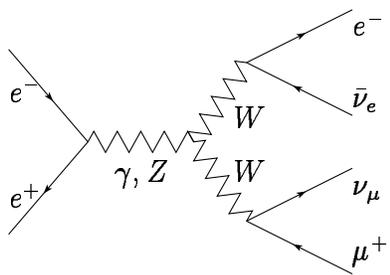
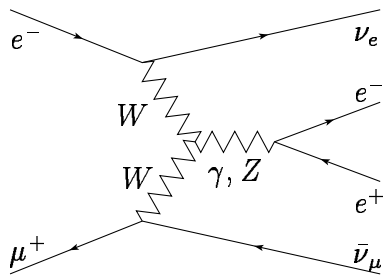
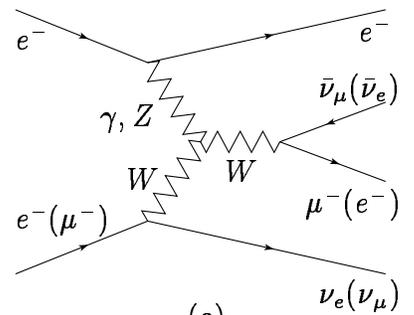
\end{document}